\documentclass[conference]{IEEEtran}
\IEEEoverridecommandlockouts
\usepackage{cite}
\usepackage{amsmath,amssymb,amsfonts}
\usepackage{algorithmic}
\usepackage{graphicx}
\usepackage{textcomp}
\usepackage{xcolor}
\usepackage{url}
\usepackage{array}
\usepackage{enumitem}
\usepackage{listings}
\usepackage{tcolorbox}
\usepackage{booktabs}

\lstdefinelanguage{SysML}{
  morekeywords={
    package, import, part, def, attribute, redefines,
    port, connect, to
  },
  sensitive=true,
  morecomment=[l]{//},
  morestring=[b]"
}

\lstdefinestyle{modelstyle}{
  basicstyle=\ttfamily\scriptsize,
  breaklines=true,
  breakatwhitespace=false,
  numbers=left,
  numberstyle=\tiny,
  stepnumber=1,
  numbersep=5pt,
  frame=single,
  columns=fullflexible,
  keepspaces=true,
  showstringspaces=false,
  captionpos=b
}

\tcbuselibrary{skins,breakable}

\newcommand{\promptfontsize}{\footnotesize}

\newtcolorbox{promptbox}[1][]{
  breakable,
  enhanced,
  colback=gray!5,
  colframe=gray!40,
  fontupper=\ttfamily\promptfontsize,   
  fonttitle=\bfseries\promptfontsize,
  title=#1,
  left=1em, right=1em, top=1em, bottom=1em,
  sharp corners
}

\begin{document}

\title{LLM-Driven Approach to Modeling Tool Interoperability in Automotive Domain}

\author{
\IEEEauthorblockN{
Nenad Petrovic, Jiajie Zhang, Vahid Zolfaghari and Alois Knoll
}

\IEEEauthorblockA{
\textit{Chair of Robotics, Artificial Intelligence and Real-Time Systems}\\
Technical University of Munich, Munich, Germany\\
Email: \{nenad.petrovic, jiajie.zhang, v.zolfaghari, k\}@tum.de
}

\thanks{
This work has received funding from the European Chips Joint Undertaking under Framework Partnership Agreement No.~101139789 (HAL4SDV), including national funding from the Federal Ministry of Research, Technology and Space of Germany under grant number 16MEE00471K. The responsibility for the content of this publication lies with the authors.
}
}

\maketitle

\begin{abstract}
Interoperability between heterogeneous modeling tools remains a significant challenge in Model-Driven Engineering (MDE), particularly in the automotive domain where multiple modeling languages, as well as defacto standard proprietary and open-source tools coexist. This paper presents an LLM-driven approach for automated model interoperability by considering two relevant aspects: 1) mapping model instances to a target metamodel 2) merging of metamodels. The proposed methodology is demonstrated through transformations involving Ecore and SysML v2 based metamodels and incorporates structural validation of generated model instances against user-defined target models. Automotive case studies illustrate the feasibility of the approach and show that large language models can significantly reduce manual transformation effort while generating structurally valid target models for cross-tool interoperability.

\end{abstract}


\section{Introduction}

Model-Driven Engineering (MDE) has become a fundamental engineering paradigm for the development of modern automotive systems, particularly in the era of Software-Defined Vehicles (SDVs), where complex software, hardware, and system architectures are developed using a variety of modeling languages and engineering tools \cite{macher2015toolchain}. Industrial environments, including CATIA Magic \cite{catiamagic} and IBM Engineering Systems Design Rhapsody \cite{rhapsody}, provide comprehensive support for standards such as SysML \cite{sysml2}, while Ecore-based ecosystems \cite{ecore}, including the Eclipse Modeling Framework (EMF), are widely used for domain-specific modeling and model-driven software engineering. As organizations increasingly employ heterogeneous modeling platforms across the development lifecycle, interoperability between these tools has become a major challenge. Engineering artifacts are frequently exchanged between tools based on different metamodels, requiring model transformations that preserve semantics while conforming to the target modeling language.

Conventional model transformation techniques rely on manually developed transformation rules, such as ATLAS Transofrmation Language (ATL) \cite{atl} or Query/View/Transformation (QVT) \cite{qvt} mappings, which require extensive domain expertise and detailed knowledge of both source and target metamodels. Developing and maintaining these transformations is labor-intensive, particularly as metamodels evolve or when multiple target modeling environments must be supported \cite{macher2015toolchain} . Moreover, transformations between heterogeneous modeling technologies often require semantic interpretation beyond straightforward structural mappings, making rule-based approaches increasingly difficult to scale.

Recent advances in Large Language Models (LLMs) have demonstrated remarkable capabilities when it comes to text summarization, understanding structured and unstructured textual artifacts, reasoning over domain concepts, and generating formal representations. These capabilities suggest that LLMs can serve as intelligent transformation engines capable of automatically generating target model instances while preserving the semantics of the source model. However, ensuring that generated models conform to the target metamodel remains a critical challenge, necessitating automated validation and refinement mechanisms.

This paper presents an LLM-driven approach for modeling tool interoperability that automatically transforms model instances between heterogeneous modeling technologies and merges multiple metamodels into a unified metamodel in a specified target format. Rather than generating explicit transformation rules, the proposed approach directly produces target-compliant model instances for model transformation tasks and unified target-compliant metamodels for metamodel integration tasks. To ensure correctness, all generated artifacts are structurally validated against the corresponding user-defined target metamodel or target metamodel specification, enabling the automatic detection of inconsistencies and supporting iterative refinement and regeneration when necessary.

The proposed methodology is demonstrated using representative automotive case studies covering four transformation scenarios: 1) Ecore-to-Ecore transformations between different metamodels, 2) SysML v2-to-SysML v2 transformations, 3) Ecore-to-SysML v2 transformations, and 4) SysML v2-to-Ecore transformations. These scenarios reflect common interoperability requirements encountered in modern automotive engineering environments involving heterogeneous modeling tools. The goal of the proposed approach is to demonstrate how generative AI can significantly simplify model interoperability workflows, reduce manual engineering effort, and facilitate the integration of heterogeneous modeling ecosystems in future automotive MDE environments.


\section{Related Works}
Semantic interoperability has become one of the MDE bottlenecks, particularly with the emergence of heterogeneous engineering toolchains and digital engineering ecosystems. While traditional interoperability approaches rely on standardized exchange formats and rule-based model transformations, recent research has investigated the use of Large Language Models (LLMs) to infer semantic correspondences between independently developed models. 

Li \emph{et al.}~\cite{li2025semantic} proposed an LLM-assisted framework for semantic alignment and integration of collaborative SysML v2 models. Their approach combines prompt engineering with iterative semantic matching and verification to identify correspondences between independently developed models while exploiting SysML v2 constructs such as aliases, imports, and metadata to preserve traceability. The work demonstrates that LLMs can effectively support semantic interoperability without requiring manually defined transformation rules.

Bader \emph{et al.}~\cite{bader2025grag} investigated interoperability from a knowledge-centric perspective by integrating SysML v2 repositories into a Graph Retrieval-Augmented Generation (GRAG) pipeline. Their approach represents SysML v2 models as semantic knowledge graphs, enabling relevant model fragments to be retrieved and injected into LLM prompts. The resulting framework improves natural-language interaction with engineering models while preserving the underlying semantic relationships between model elements.

Recent work has also explored AI-assisted generation and integration of SysML models. In ~\cite{hendricks2025text2model} , the authors oproposed an end-to-end workflow that automatically derives SysML diagrams from natural-language documents before generating executable computational models. Their framework combines NLP techniques and LLMs to extract entities, relationships, and block definitions, demonstrating the role of semantic modeling as an intermediate representation for interoperability.

Although these approaches significantly advance semantic interoperability within the SysML ecosystem, they primarily focus on semantic alignment, knowledge retrieval, or model generation. In contrast, the proposed framework addresses the complementary problem of instance model interoperability across heterogeneous metamodels. Given an existing model instance and an arbitrary target metamodel, the framework employs prompt-guided LLM reasoning to perform semantic migration and subsequently validates the generated instance against the target metamodel. By integrating conversational orchestration, semantic mapping, and automated conformance validation into a unified workflow, the proposed approach enables end-to-end interoperability between heterogeneous model-driven engineering environments.

The proposed approach builds upon our previous experiences in LLM-driven Ecore metamodel construction \cite{petrovic2025metamodeling}, model instance creation and validation \cite{petrovic2025genai}. For implementation, we adopt good practices of leveraging n8n workflow \cite{n8n2026}. Additionally, in this work we primarily focus on adoption of locally deployable LLMs, as control over information flow is one of main barriers when it comes to AI-based automation in critical areas, such as automotive \cite{petrovic2025survey}.

\section{Implementation overview}

\begin{figure}[t]
    \centering
    \includegraphics[width=\columnwidth]{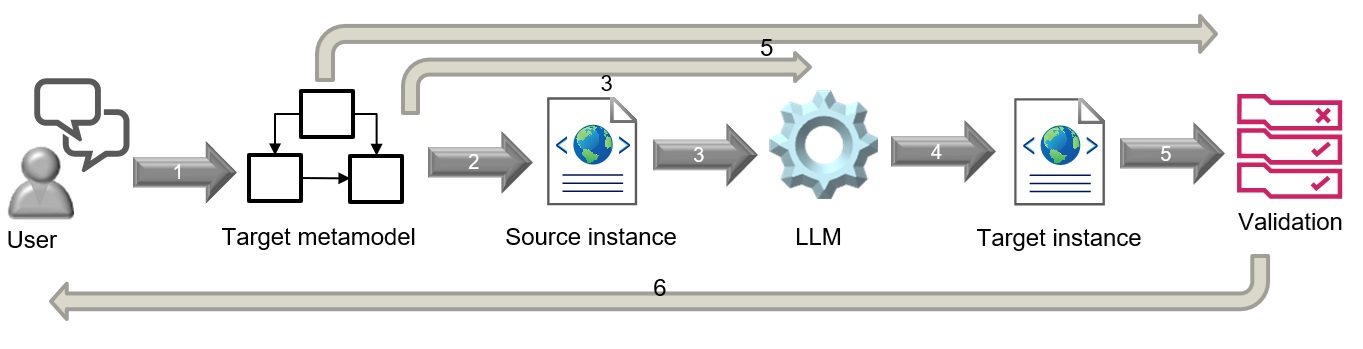}
    \caption{Integrated code generation toolchain based on n8n.}
    \label{fig:workflow}
\end{figure}

\begin{figure}[t]
    \centering
    \includegraphics[width=\columnwidth]{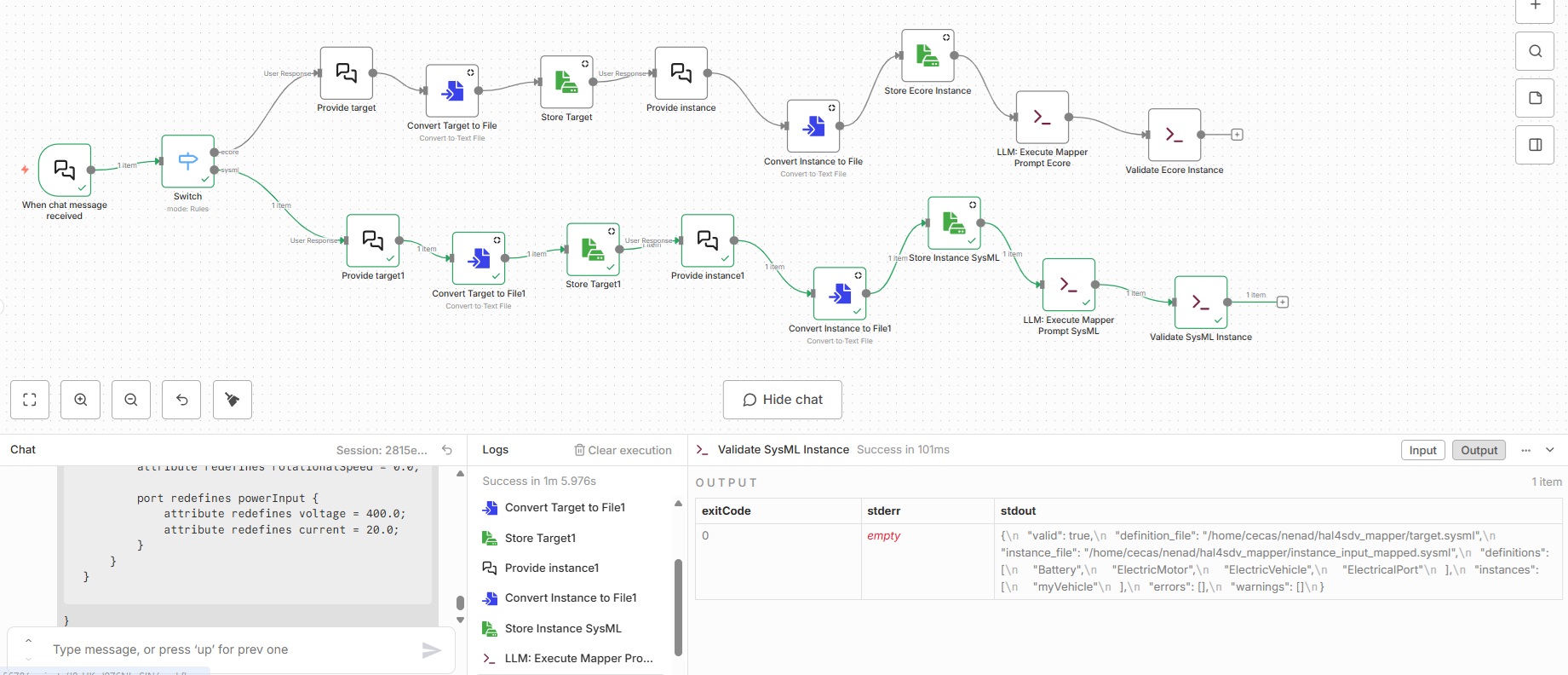}
    \caption{Integrated code generation toolchain based on n8n.}
    \label{fig:n8n_mapper}
\end{figure}

Automated workflow was implemented using \textit{n8n}, integrating conversational interaction, file management, LLM inference, and model validation into a single execution pipeline. Conceptual workflow, shown in Fig.~\ref{fig:workflow}, supports both \textit{Ecore} and \textit{SysML} mapping scenarios, while the screenshot of its n8n implementation can be seen in Fig. \ref{fig:n8n_mapper} .

The execution begins when a user submits a request through the chat interface. A routing component determines the desired mapping variant (Ecore or SysML) and activates the corresponding processing branch. The workflow subsequently requests two artifacts from the user: 1) the target metamodel and 2) the source model instance. Uploaded files are converted into plain-text representations and stored locally for subsequent processing.

The mapping stage (denoted as step 3) invokes a Python-based orchestration script that reads both the source instance model and the target metamodel and constructs the LLM prompt dynamically. Two prompts are employed. The system prompt specializes the LLM as a model transformation engine:

\begin{quote}
\small
\textit{You are converting a given model instance to comply with a target Ecore metamodel. Map as many elements as possible from the model instance to the target metamodel. The result must be an Ecore XMI model instance compliant with the target metamodel.}
\end{quote}

On other side, for SysML, the system prompt requires additional explicit instructions in order to avoid ambiguity coming from the nature of the notation itself.
\begin{quote}
\small
\textit{You are an expert in SysML v2 model transformation. Convert the given source model instance into a SysML v2 textual instance that conforms to the provided target SysML model definition. Preserve as much semantic information as possible while ensuring that the generated model conforms to the target definitions. Generate the output as a SysML v2 instance using the same structure as the target model, including a package declaration, required imports, a typed top-level part usage, nested ``part redefines'' statements, and ``attribute redefines'' assignments for property values. Create a package containing typed top-level part usages using the required SysML v2 syntax. Reuse only definitions declared in the target SysML model. Return only the resulting SysML v2 model without explanations, comments, or Markdown code fences.}
\end{quote}

The corresponding user prompt embeds the complete contents of both uploaded models:

\begin{quote}
\small
\textit{Map the following model instance: \{instance model contents\} to comply with the following target metamodel: \{target metamodel contents\}. Return only the resulting model instance without explanations or Markdown code fences.}
\end{quote}

The mapping request is submitted to an OpenAI-compatible inference endpoint hosting the LLMs. The generated response is post-processed to remove potential Markdown formatting before being stored as output file representing the target instance model (step 4).

To ensure correctness of the generated model, the workflow automatically executes a validation stage following the mapping process (step 5). For Ecore-based transformations, a dedicated Python validator, implemented using the PyEcore framework, loads both the target Ecore metamodel and the generated XMI instance to verify structural and semantic conformance. The validation process performs XML parsing, resolves the metamodel, instantiates the model according to the corresponding Ecore definitions, and evaluates EMF validation constraints. The performed checks include XML well-formedness, metamodel conformance, containment hierarchy consistency, reference resolution, multiplicity constraints, attribute type compatibility, and mandatory feature validation. All detected inconsistencies are classified as errors or warnings and returned to the workflow, enabling immediate feedback on the quality of the generated model. For SysML-based transformations, the workflow follows the same orchestration and LLM mapping pipeline, while the validation component can be replaced by a SysML-specific conformance checker as supported by the underlying modeling environment. This modular architecture separates model generation from validation, allowing additional modeling languages and domain-specific metamodels to be integrated with minimal changes to the overall workflow. Integrating automated validation into the transformation pipeline provides a formal verification step that complements the generative capabilities of the LLM and supports iterative refinement by identifying precisely which model elements violate the target metamodel.
The proposed workflow combines conversational interaction, automated file handling, LLM-driven semantic transformation, and formal model validation into a reproducible pipeline. Integrating validation directly into the generation process enables rapid feedback regarding the correctness of generated instance models and provides a foundation for iterative refinement of LLM-based model transformations (step 6), using the update prompt in a similar way as in our previous work \cite{petrovic2025metamodeling}:

\begin{quote}
\small
\textit{Update the model instance: \{instance model contents\} based on validation outcomes: \{validation results\}. }
\end{quote}

In a similar manner, a workflow is defined for merging two metamodels into a selected target format. The objective is to generate a unified metamodel that preserves all unique concepts, relationships, and constraints from both source metamodels while ensuring compliance with the target representation. The underlying prompt follows the structure shown below:

\begin{quote}
\small
\textit{Merge the following two source metamodels into one unified target metamodel. First source metamodel: \{first source metamodel contents\}. Second source metamodel: \{second source metamodel contents\}. Target modeling language: \{target modeling language\}. Preserve all compatible classes, attributes, references, inheritance relationships, multiplicities, containment relationships, and data types. Identify equivalent concepts and merge them instead of creating duplicates. Integrate complementary concepts from both metamodels. Resolve naming and structural conflicts consistently. Preserve the semantics of both source metamodels as much as possible. Generate one valid metamodel in the requested target modeling language. Return only the resulting metamodel without explanations, comments, or Markdown code fences.}
\end{quote}
For metamodel merging, we rely on workflow based on the one presented in \ref{fig:workflow}, with two main differences: 1) instead of instance model, in the second step we provide a metamodel 2) validation of metamodels, instead of instances.

\section{Automotive Hardware Abstraction Scenarios}
\subsection{Model Instance Mapping}
To demonstrate the proposed framework, an illustrative automotive hardware abstraction scenario is considered in which an existing hardware configuration represented as a SysML v2 instance model is automatically transformed into an Ecore/XMI instance conforming to a user-defined Ecore metamodel. 
In the considered example, the source SysML v2 instance model (given in Listing \ref{lst:sysml-source}) describes a sensor platform comprising a front-facing camera and a LiDAR sensor. The model captures the hardware architecture through typed part usages and redefined attributes, including sensor resolution, field of view, and sensing range. 

\begin{lstlisting}[
  language=SysML,
  style=modelstyle,
  caption={Source SysML v2 instance model.},
  label={lst:sysml-source}
]
package AutomotiveHardwareInstances {

    import AutomotiveHardwareDefinitions::*;

    part frontSensorPlatform : SensorPlatform {

        part redefines camera {
            attribute redefines resolution = "1920x1080";
            attribute redefines fieldOfView = 120.0;
        }

        part redefines lidar {
            attribute redefines resolution = "0.2";
            attribute redefines fieldOfView = 360.0;
            attribute redefines range = 200.0;
        }
    }
}
\end{lstlisting}

\begin{figure}[t]
    \centering
    \includegraphics[width=0.6\columnwidth]{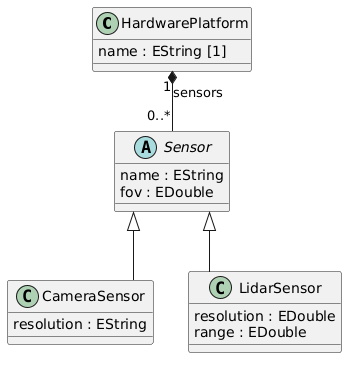}
    \caption{Target Ecore metamodel.}
    \label{fig:target_ecore}
\end{figure}

The target Ecore metamodel (shown in Fig.\ref{fig:target_ecore}) represents the same hardware configuration using Ecore classes, containment references, and typed attributes, providing a machine-readable representation suitable for EMF-based applications.

The objective of the transformation is to preserve the semantics of the original hardware configuration while expressing it as an Ecore-compliant XMI instance. Instead of relying on manually specified transformation rules, the proposed framework provides both the source SysML v2 instance model and the target Ecore metamodel directly to the LLM. Based on the semantic descriptions contained in both artifacts, the model infers the correspondence between SysML part definitions and Ecore classes (e.g., \textit{Camera} $\rightarrow$ \textit{CameraSensor}, \textit{Lidar} $\rightarrow$ \textit{LidarSensor}) and maps redefined properties such as resolution, field of view, and sensing range to the corresponding Ecore attributes. The generated XMI instance (shown in Listing \ref{lst:xmi-generated}) preserves the hierarchical structure of the original system while conforming to the structural constraints imposed by the target metamodel.

\begin{lstlisting}[
  language=XML,
  style=modelstyle,
  caption={Generated XMI instance conforming to the target Ecore metamodel.},
  label={lst:xmi-generated}
]
<?xml version="1.0" encoding="UTF-8"?>
<hw:HardwarePlatform
    xmi:version="2.0"
    xmlns:xmi="http://www.omg.org/XMI"
    xmlns:xsi="http://www.w3.org/2001/XMLSchema-instance"
    xmlns:hw="http://example.org/automotive/hardware"
    name="frontSensorPlatform">

    <sensors
        xsi:type="hw:CameraSensor"
        name="camera"
        resolution="1920x1080"
        fov="120.0"/>

    <sensors
        xsi:type="hw:LidarSensor"
        name="lidar"
        resolution="0.2"
        fov="360.0"
        range="200.0"/>

</hw:HardwarePlatform>



\end{lstlisting}

\subsection{Metamodel Merging}

To demonstrate the metamodel merging capability, let us assume that we want to merge the SysML v2 metamodel describing vehicle actuators, shown in Listing~\ref{lst:sysml-source-merge}, with the previously introduced Ecore automotive hardware metamodel from Fig. ~\ref{fig:target_ecore}. The Ecore metamodel already defines the platform and sensor hierarchy, while the SysML model introduces the actuator hierarchy together with a vehicle control component containing brake and steering actuators. 
\begin{lstlisting}[
  language=SysML,
  style=modelstyle,
  caption={Source SysML v2 metamodel covering actuator aspects used for merging with the Ecore automotive hardware metamodel.},
  label={lst:sysml-source-merge}
]
package VehicleDefinitions {

    abstract part def Actuator {
        attribute name : String;
    }

    part def BrakeActuator
        specializes Actuator {

        attribute brakingLevel : Real;
    }

    part def SteeringActuator
        specializes Actuator {

        attribute steeringAngle : Real;
    }

    part def VehicleControl {

        part brakes : BrakeActuator;
        part steering : SteeringActuator;
    }
}
\end{lstlisting}

\begin{figure}[t]
    \centering
    \includegraphics[width=0.9\columnwidth]{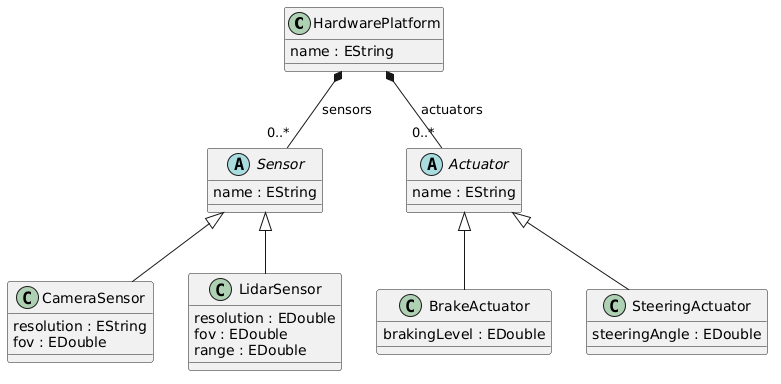}
    \caption{Merged Ecore metamodel containing both sensors and actuators.}
    \label{fig:merged_ecore}
\end{figure}

The objective of the merge operation is to combine the complementary concepts from both metamodels into a single unified Ecore metamodel without introducing duplicate or conflicting elements. During the transformation, equivalent concepts are identified, inheritance hierarchies are preserved, and the actuator definitions together with their attributes and containment relationships are integrated into the target metamodel. The resulting merged metamodel contains both sensing and actuation concepts within a common automotive hardware representation, enabling subsequent model instances to simultaneously describe sensors and actuators while remaining compliant with a single Ecore metamodel. The result is depicted in Fig. \ref{fig:merged_ecore}.

\section{Evaluation}
The proposed LLM-driven model mapping framework was evaluated on four representative transformation scenarios: \textit{Ecore-to-Ecore}, \textit{SysML v2-to-SysML v2}, \textit{Ecore-to-SysML v2}, and \textit{SysML v2-to-Ecore}. The evaluation compares two locally deployable LLMs, \texttt{google/gemma-4-31B-it} and \texttt{Qwen/Qwen3.5-122B-A10B}, on their ability to automatically generate target model instances that conform to the corresponding target metamodel. Both homogeneous and heterogeneous transformations were considered to assess the generality of the proposed approach. The models are deployed locally using an OpenAI API-compatible inference server running on three NVIDIA H200 NVL GPUs (141 GB each), while default model parameters available in documentation were used \cite{gemma4}\cite{qwen35}.

A benchmark is based on representative automotive hardware models describing sensing and actuation of a vehicular system. The target metamodel consists of 27 elements, covering camera, LiDAR, radar, brake, and steering components. The corresponding source instance models contains hierarchical compositions with typed attributes and engineering parameters representative in automotive area. It contains 1 hardware platform, 5 component instances (3 sensors and 2 actuators), 16 attribute-value assignments, and 5 containment relationships. Reference transformations were manually prepared with expert support and used as the ground truth for evaluation. Two aspects of result quality were considered: 1) semantic matching - percentage of semantically equivalent elements mapped 2) syntax correctness - percentage of tries (out of 5) when the generated instance model passes the corresponding validation script.

Table~\ref{tab:mapping_results} summarizes the obtained results. The last two rows consider average performance and execution time based on 5 runs for each of the models. Both evaluated LLMs achieved complete semantic mapping across all four transformation scenarios, correctly preserving every expected model element regardless of whether the transformation was performed between homogeneous (Ecore$\rightarrow$Ecore and SysML v2$\rightarrow$SysML v2) or heterogeneous (Ecore$\rightarrow$SysML v2 and SysML v2$\rightarrow$Ecore) modeling languages. This demonstrates that current locally deployable LLMs are capable of inferring semantic correspondences between heterogeneous metamodels without relying on manually specified transformation rules.

\begin{table}[t]
\caption{Comparison of model instance mapping performance across transformation scenarios (semantic matching/syntax correctness percentage).}
\label{tab:mapping_results}
\centering
\footnotesize
\begin{tabular}{lcc}
\toprule
\textbf{Scenario} & \textbf{Gemma} & \textbf{Qwen} \\
\midrule
Ecore $\rightarrow$ Ecore &
100\% / 100\% &
100\% / 100\% \\

Ecore $\rightarrow$ SysML v2 &
100\% / 80\% &
100\% / 60\% \\

SysML v2 $\rightarrow$ SysML v2 &
100\% / 60\% &
100\% / 40\% \\

SysML v2 $\rightarrow$ Ecore &
100\% / 100\% &
100\% / 100\% \\
\midrule
Average performance &
100\% / 85\% &
100\% / 75\% \\

Time (s) &
\textbf{5.1} &
46.5 \\
\bottomrule
\end{tabular}
\end{table}

The primary differences between the evaluated models are reflected in the syntactic correctness of the generated models. For transformations targeting Ecore/XMI, both models consistently generated syntactically valid instances that successfully passed the automated validation stage. In contrast, SysML v2 generation proved considerably more challenging. While the generated models remained semantically complete, both LLMs occasionally produced SysML v2 syntax that deviated from the expected textual notation, primarily through incorrect use of \texttt{redefines} constructs, inconsistent import statements, or generation of typed part usages instead of inherited feature redefinitions. Consequently, syntax correctness decreased to 80\% and 60\% for the Ecore$\rightarrow$SysML v2 scenario and to 60\% and 40\% for SysML v2$\rightarrow$SysML v2 for Gemma and Qwen, respectively.

These results indicate that the principal limitation of current LLM-based model transformation lies not in semantic understanding, but in strict adherence to the formal textual syntax required by SysML v2. Since the generated models preserve the complete engineering semantics, many of these syntactic inconsistencies could be potentially addressed through more specific prompts or lightweight post-processing.

The generated instance models were automatically validated against the target metamodel using dedicated Ecore and SysML v2 validators. Validation detects structural inconsistencies, missing mandatory elements, incorrect references, typing violations, and serialization errors, ensuring that only metamodel-compliant models are accepted. The average end-to-end execution time, including model generation and validation, was approximately 5.1~s for \texttt{google/gemma-4-31B-it} and 46.5~s for \texttt{Qwen/Qwen3.5-122B-A10B}. The lower execution time of Gemma is primarily attributed to its smaller computational footprint and higher inference throughput, whereas Qwen's larger architecture incurs additional reasoning and generation overhead. All reported results represent the average of five independent executions.

While For the SysML v2$\rightarrow$Ecore transformation, both Gemma and Qwen achieved 100\% semantic mapping accuracy while producing syntactically valid XMI instances, in few cases - Gemma generated non-standard XMI serialization (e.g., qualified containment elements). However, the semantic content remained complete and only minor serialization-related corrections were required to obtain a fully compliant model.

On the other side, for model merging, the experiment combines two complementary metamodels while excluding intermediate container elements from the evaluation. The first metamodel contains four relevant classes describing sensors: \texttt{Sensor}, \texttt{CameraSensor}, \texttt{LidarSensor}, and \texttt{RadarSensor}. It defines five domain-relevant attributes. The second metamodel contains three relevant classes: \texttt{CameraSensor}, \texttt{BrakeActuator}, and \texttt{SteeringActuator}, with seven attribute definitions.

The \texttt{CameraSensor} class is present in both metamodels and represents the main overlapping concept. Its attributes \texttt{name}, \texttt{resolution}, and \texttt{fov} also overlap semantically. The remaining three classes from the first one and remaining two classes from the second are complementary, non-overlapping concepts. Consequently, the expected merged metamodel contains six unique domain classes and nine unique attribute definitions. A correct merge should preserve the shared camera concept only once while integrating the lidar, radar, brake, and steering concepts without information loss. The results of model merging performed under the same conditions and summarized in Table \ref{tab:merge_results}. When it comes to both the result matching and average execution time, similar trends can be also noticed on metamodel-level merging in this case, with Gemma's slight advantage regarding the syntactical correctness in case of SysML target format.

\begin{table}[t]
\caption{Comparison of metamodel merging performance across target modeling languages (semantic matching/syntax correctness percentage).}
\label{tab:merge_results}
\centering
\footnotesize
\begin{tabular}{lcc}
\toprule
\textbf{Scenario} & \textbf{Gemma} & \textbf{Qwen} \\
\midrule
Ecore + SysML v2 $\rightarrow$ Ecore &
100\% / 100\% &
100\% / 100\% \\

Ecore + SysML v2 $\rightarrow$ SysML v2 &
100\% / 60\% &
100\% / 40\% \\
\midrule
Average performance &
100\% / 80\% &
100\% / 70\% \\

Time (s) &
\textbf{15.2} &
58.3 \\
\bottomrule
\end{tabular}
\end{table}


The complete implementation of the proposed framework, including the model mapping and merging workflows, Ecore and SysML v2 validators, example metamodels, model instances, and all experimental scenarios used in this paper, is publicly available at \url{https://github.com/np-tum-air/tum_hal4sdv_model_mapper/}.

\section{Conclusion}

This paper presented an LLM-assisted workflow for semantic instance model mapping across heterogeneous metamodels, combining conversational, prompt-driven transformation and automated model validation into a unified pipeline. By leveraging locally deployable LLMs to infer semantic correspondences directly from the source instance model and target metamodel, the proposed approach eliminates the need for manually specified transformation rules while maintaining structural conformance through formal validation. The integration of n8n-based orchestration enables an end-to-end automated process encompassing artifact acquisition, model transformation, validation, and result reporting.

Compared with traditional rule-based transformation techniques, the proposed framework offers greater flexibility when dealing with evolving metamodels, heterogeneous modeling languages, and domains where explicit transformation rules are unavailable or prohibitively expensive to develop and maintain. The modular architecture allows the same orchestration pipeline to support multiple modeling environments, currently including Ecore and SysML, by adapting prompts and validation components rather than redesigning transformation logic. This substantially lowers the effort required to establish interoperability between modeling ecosystems while preserving compatibility with existing MDE toolchains.

When it comes to future works and further research directions, we will work on handling large model files using Model Context Protocol (MCP) and Retrieval Augmented Generation (RAG) in order to tackle the cases when context size might not be large enough to include the full model, building upon our previous work from \cite{mazur2025agentic} .


\vspace{12pt}

\end{document}